\documentclass[aps,12pt,
superscriptaddress,amsmath,amssymb,nofootinbib,preprintnumbers]{revtex4-1}

\usepackage[breaklinks = true,colorlinks = true,%
linkcolor = blue,citecolor = blue]{hyperref}
\usepackage[dvipdfmx]{graphicx}
\usepackage{xcolor}
\usepackage{comment}

\newcommand{\cG}{{\cal G}}
\newcommand{\cO}{{\cal O}}
\newcommand{\bbZ}{{\mathbb Z}}

\newcommand{\be}{\begin{eqnarray}}
\newcommand{\ee}{\end{eqnarray}}

\newcommand{\SL}{\mathrm{SL}}
\newcommand{\nt}{\notag\\}

\begin{document}

\title{Flavor Moonshine}

\author{Shotaro Shiba Funai}
\email{shotaro.funai@oist.jp}
\affiliation{Physics and Biology Unit, Okinawa Institute of Science and Technology (OIST), 
 1919-1 Tancha, Onna-son, Kunigami-gun, Okinawa 904-0495, Japan}

\author{Hirotaka Sugawara}
\email{sugawara@post.kek.jp}
\affiliation{High Energy Accelerator Research Organization (KEK),
 1-1 Oho, Tsukuba, Ibaraki 305-0801, Japan}

\begin{abstract}
\vspace*{10pt}
The flavor moonshine hypothesis 
is formulated to suppose 
that all particle masses (leptons, quarks, Higgs and gauge particles -- more precisely, their mass ratios) are expressed as coefficients in the Fourier expansion of some modular forms just as, in mathematics, dimensions of representations of a certain group are expressed as coefficients in the Fourier expansion of some modular forms. 
The mysterious hierarchical structure of  the quark and lepton masses is thus attributed to that of the Fourier coefficient matrices of certain modular forms.
{Our intention here is} 
not to {prove this hypothesis} 
starting from some physical assumptions 
but {rather} to demonstrate 
that this hypothesis is experimentally verified and, 
assuming that the string theory correctly describes the natural law, 
to calculate the geometry (K\"{a}hler potential and the metric) of the moduli space of the Calabi-Yau manifold, thus providing a way to calculate the metric of Calabi-Yau manifold itself directly from the experimental data.
\end{abstract}

\preprint{KEK-TH-2151}

\maketitle

\section{Introduction}

Some researchers including one of the authors of this work (H.\,S.) have been working on flavor physics, assuming that some discrete symmetry plays an important role in its understanding~\cite{ref1};
$S_3, S_4, A_4$, etc.
But the outcome is very limited and so far we have no clear understanding of flavor physics. 
Topological definition of Higgs Yukawa coupling also has not led to any useful prediction on the flavor physics {to date}~\cite{ref6}. 

On the mathematical side, 
a dramatic phenomenon called ``moonshine" {has been described}~\cite{ref2}, 
in which a discrete symmetry (specifically, dimensions of representation of the monster group) is manifested in a modular form in a rather unexpected manner. 
When this happens, 
we may use this fact for the discrete symmetry in flavor physics: We start by assuming that the symmetry of flavor physics is manifested in a certain modular form. 
Corresponding to each flavor we assume {such a} 
modular form. 
The modular forms must contain all the information 
{about} flavor physics with the understanding that all this 
information 
is contained in the Higgs coupling to leptons and hadrons. 

More precisely, we assume that the particle masses (the mass ratios), rather than dimensions of a representation of discrete group, are directly written in the Fourier coefficients of these modular forms --- the flavor moonshine hypothesis. 
The mass ratios are scale independent quantities~\cite{ref3} and do not 
{vary with} energy scale. 
We observe that at least in the lowest order perturbation calculations 
the logarithmic scale dependence 
cancels out completely both in QCD and in EW theory,
although it does not exclude the renormalization effect proportional to such terms as 
$\log(m_1/m_2)$.
We refer to the reference~\cite{ref3} here for the non-perturbative calculations.
Therefore, the mass ratio is an appropriate quantity to discuss physics even at the highest energy scale. The gauge particle masses must also be written as some modular forms but we will not discuss 
{that matter} in this work. 

In pure mathematics, we anticipate generalization of conventional ``moonshine" from single variable modular form to multi-variable modular form. 
Certain mathematical ``object,'' 
perhaps the representation matrix of a certain group
rather than 
the dimension of the monster group, must be written in the Fourier expansion of the multi-variable modular forms.
We will identify the mass matrix {with} this ``object.''

The question {arises}: 
What are those modular forms 
{that} manifest the discrete symmetry 
{appearing} in flavor physics? 
For the time being, we postpone the question of justifying our 
{adoption of a} certain modular form to each flavor based on a general formalism such as string theory, but rather we proceed backward and 
investigate {instead} what the experimentally acceptable modular forms are. 
We then 
determine what kind of geometry can 
{yield} such a modular form when we consider the compactification of the string theory. 

We define the flavor modular form in the following way.
Suppose we have a two variable modular form to each flavor. 
Then it can be Fourier expanded as
\be \label{eq1}
J(q,r) = \frac{1}{g}\sum_{i=0, j=-\infty}^\infty g_{ij} q^i r^j
\ee
where $g_{ij}$ for $i\geq 0$ and $g_{i,-j}=g_{ij}$ for the symmetric modular form~\cite{ref4}.
The $g_{ij}$ is supposed to correspond to the Higgs coupling of $i$ and $j$ quarks or to the corresponding leptons. 
By solving the equation (\ref{eq1}) backwards we have
\be
g_{ij} = g\int_0^1 \int_0^1 J(q,r) q^{-i} r^{-j} d\tau d\sigma
= \frac{g}{(2\pi i)^2} \int_C \int_C J(q,r) q^{-i-1} r^{-j-1} dq dr.
\ee
Here, $q=e^{2\pi i\tau}$ and $r=e^{2\pi i\sigma}$.
The integration is done along the circle $C$ of radius 1 with the center at the origin. 
It is important that we 
{integrate} over the modular variables to 
{obtain} the coefficient. 

If the modular form is based on the ring of integers, the forms are numerous and it is hard to pinpoint the appropriate form.
Fortunately, if we generalize the integer ring appropriately 
{to constrain} the possible forms, 
{then} in the case we are considering where $g_{i,-j}=g_{ij}$, called the symmetric modular form, it is known that all the modular forms can be constructed rather easily~\cite{ref4}.

Specifically, as the simplest generalization, we use $\SL(2,\bbZ(\sqrt{2}))$ to define the flavor modular group of the two variable modular form rather than $\SL(2,\bbZ)$.\footnote{
When we thought of flavor moonshine, it was clear that the relevant modular form must have more than one variables. 
It also seemed the $\SL(2,\bbZ)$ is 
{insufficiently constrained, allowing} too many choices for the forms. 
Therefore, we looked for some work enlarging $\SL(2,\bbZ)$ so that the choice becomes manageable, and we encountered a paper by H. Cohn and J. Deutsch~\cite{ref4} 
{where we} learned that there are only three generators for the entire $\SL(2,\bbZ(\sqrt{2}))$, which is the simplest kind of $\SL(2,\bbZ)$ extension. 
To our great surprise, we found its modular form in the lowest level ($k=1$) describes the charged lepton mass ratios correctly (in section \ref{sec:2.1}).}
We put
\be
q = e^{\pi i(z+z')},\quad
r=e^{\pi i(z-z')/\sqrt{2}};\quad
2\tau = z+z',\quad
2\sqrt{2}\sigma=z-z'.
\ee
Then the condition for the modularity is the transformation property:
\be
J(e^{2\pi iz}, e^{2\pi iz'})
~\to~
J(e^{2\pi iz}, e^{2\pi iz'})\left((\gamma z+\delta)(\gamma'z'+\delta')\right)^{2k}
\ee
under
\be \label{eq4}
(z,z')\to\left(\frac{\alpha z+\beta}{\gamma z+\delta}, \frac{\alpha' z'+\beta'}{\gamma' z'+\delta'}\right), \quad
\alpha,\beta,\gamma,\delta,\ldots \in \bbZ(\sqrt{2})
\ee
and $\alpha = a+b\sqrt{2}, \alpha' = a-b\sqrt{2}, \ldots$
with $a,b:$ integer.
$2k$ is called the ``level''.

Cohen-Deutsch~\cite{ref4} shows that there are only three generator modular forms in this case. 
They are given by $G_2,G_4,G_6$ with $k=1, 2, 3$. 
What we use are the coefficients in Fourier expansion of these modular forms.
We may also choose different combinations $H_2,H_4,H_6$ 
{that} are given by
\be
H_2=G_2,\quad
H_4=\frac{11G_2^2-G_4}{576},\quad
H_6=\frac{361G_2^3-G_6-50976G_2H_4}{224640}.
\ee

We have only one modular form for $k=1$: $G_2$,
two forms for $k=2$: $G_4$ and $G_2^2$,
and three forms for $k=3$: $G_6, G_2^3$ and $G_2G_4$.
Linear combination of {forms of the} same level $2k$ 
is again a modular form. 
Therefore all modular forms up to the level 6 ($k=3$) are given by
\be
k=1 &:&\quad G_2 \label{eq6}\\
k=2 &:&\quad G_4+a_4G_2^2 \label{eq7}\\
k=3 &:&\quad G_6+a_6G_2^3+b_6G_2G_4 \label{eq8}
\ee
where $a_4,a_6$ and $b_6$ are complex numbers.

In order to write down the Higgs coupling of quarks and leptons, we define the following:
First we define, for the Higgs coupling of a certain flavor,
\be\label{eq9}
F(q^{-1},r^{-1})\equiv gH\lim_{G\to\infty} \sum_{i,j=0}^{G-1} \overline\psi_{Rj}\psi_{Li} q^{-i} r^{-j}
\ee
where $H$ is the Higgs field and $\psi_L, \psi_R$ are quark or lepton fields.
The Yukawa coupling is given by
\be
Y&=&\int J(q,r)F(q^{-1},r^{-1})d\tau d\sigma
=\frac{1}{(2\pi i)^2}\int J(q,r)F(q^{-1},r^{-1})\frac{dqdr}{qr} \notag\\
&=&H\sum_{i,j=0}^{\infty} g_{ij}\overline\psi_{Rj}\psi_{Li}
=gH\sum_{i,j,k=0}^{\infty} U^\dagger_{Lik}\lambda_k U_{Rkj}\overline\psi_{Rj}\psi_{Li}
\ee
where
$U_L, U_R$ are unitary matrices and $\lambda$ denotes elements of diagonalized $g_{ij}$ matrix, i.e., 
\be\label{eq11}
g_{ij}=g \lim_{G\to\infty} \sum_{k=0}^{G-1} U_{Lik}^\dagger\lambda_kU_{Rkj}
\ee
for $i,j=0,1,\ldots,G-1$.
Then we have
\be\label{eq12}
Y=gH \lim_{G\to\infty}
\sum_{k=0}^{G-1} \lambda_k \overline\chi_{Rk}\chi_{Lk}
\ee
where
$\chi_{Lk}=U^\dagger_{Lik}\psi_{Li}$ and $\chi_{Rk}=\psi_{Rj}U^\dagger_{Rjk}$.
To 
{maintain} the modular invariance of the Yukawa coupling, we assume the transformation property:
\be
F(q^{-1},r^{-1})=F(z,z')
~\to~
\left((\gamma z+\delta)(\gamma' z'+\delta')\right)^{-2k+2}
F(z,z')
\ee
under the modular transformation (\ref{eq4}).
The level $-2k+2$ is to take care of the transformation property of $dq dr/qr$:
\be
\frac{dqdr}{qr} = \frac{d\tau d\sigma}{(2\pi i)^2}
=\frac{dzdz'}{2\sqrt{2}(2\pi i)^2}
~\to~
(\gamma z+\delta)^{-2} (\gamma' z'+\delta')^{-2}
\frac{dzdz'}{2\sqrt{2}(2\pi i)^2}.
\ee
If the original $\tau,\sigma$ are real, so are the transformed $\tau,\sigma$.
Therefore, the unit circle goes to the unit circle and the modular invariance is maintained.

Some remarks are in order:
\begin{enumerate}
\item
This construction suggests the definition of the fields:
\be\label{eq16}
\psi_L(x,q)=\lim_{G\to\infty}\sum_{i=0}^{G-1}\psi_{Li}q^{-i},\quad
\psi_R(x,r)=\lim_{G\to\infty}\sum_{j=0}^{G-1}\psi_{Ri}r^{-j}.
\ee
We do not need to assume any specific transformation property of the individual field under modular transformation,
while the bilinear form expressed in equation (\ref{eq9}) must transform covariantly under the modular transformation. 
We also note that the transformation property (\ref{eq9}) is consistent only when 
the number of generation $G$ is infinite. 
Finite $G$ violates the modular invariance of the Yukawa coupling.

\item
We treat here, just for simplicity, {a pristine} Higgs field $H$. 
However, in section \ref{sec:4}, 
we will define and use the modular form corresponding to the Higgs field:
\be
J_H(w)=\sum_k h_kw^k
\ee
corresponding to $J(q,r)$. We can also define the field
\be\label{eq16'}
H(w^{-1})=\sum_k H_k w^{-k}
\ee
with the Higgs field $H=H_0$.

\item
Our modular variables $q, r$ eventually become the moduli of Calabi-Yau manifold as will be shown later in section \ref{sec:4}. 
The usual treatment of these variables is to regard them as a scalar field in the four-dimensional space-time and to try to find a way to stabilize them. 
We regard them as variables to distinguish different vacua, and we integrate over them as in equation (\ref{eq12}) to 
{obtain} the Yukawa coupling. 
This roughly corresponds to superposing all possible {equivalent} vacua. 
The Yukawa interaction resolves this degeneracy, {so that} 
each value of generation $G$ corresponds to a different vacuum.
We have $G=3$ in this work 
as {it concerns} the low energy experimental data. 
It may happen that 
{a} phase transition occurs at high energy, 
{in which case} the particle masses would change suddenly at that energy scale. 

\item
Our definition of the ``generation" is not the same as the usual one in string theory. 
It corresponds to the expansion coefficient of the modulus dependent fields defined in equations (\ref{eq16}) and (\ref{eq16'}).
\end{enumerate}


\section{Numerical results}
\label{sec:2}

Equation (\ref{eq11}) 
shows that $g_{ij}$ is a mass matrix, and equation (\ref{eq1}) shows it is just the Fourier coefficient of the modular form $J(q,r)$. 
In this section we consider each case of equations (\ref{eq6}), (\ref{eq7}) and (\ref{eq8}), separately.

\subsection{Case of $k=1$}
\label{sec:2.1}

The modular form $J(q,r)=G_2$ in this case.
From a table given by Cohn and Deutsch~\cite{ref4},
we have
\be
g_{ij} = g \begin{pmatrix}
1&0&0&0&0&\cdots\\
144&48&0&0&0&\cdots\\
720&384&336&0&0&\cdots\\
1440&864&1152&480&144&\cdots\\
3024&1536&2688&1152&1488&\cdots\\
\vdots&\vdots&\vdots&\vdots&\vdots&\ddots
\end{pmatrix}.
\ee
From now on we restrict ourselves to the $G = 3$ case:
\be
g_{ij} = g \begin{pmatrix}
1&0&0\\
144&48&0\\
720&384&336
\end{pmatrix}
=: gM_3.
\ee
The mass square matrix is given by $gg^\dagger$ and it will be diagonalized as
\be\label{eq:gg}
(gg^\dagger)_{ij} = g^2U_{Lik}^\dagger |\lambda_k|^2 U_{Rkj} 
\ee
with 
sum over the indices $k$.
We diagonalize the mass square matrix $M_3^2$ and find its 
square root 
is
\be
\sqrt{M_3 M_3^T} = \begin{pmatrix}
0.2929&0&0\\
0&61.63&0\\
0&0&893.3
\end{pmatrix}.
\ee
By normalizing the lowest mass to be the electron mass of 0.5110 MeV, we 
obtain
\be
\left(\sqrt{M_3 M_3^T}\right)_\text{normalized}=\begin{pmatrix}
0.5110&0&0\\
0&107.5&0\\
0&0&1558
\end{pmatrix}.
\ee
This shows that the modular form $G_2$ 
{embodies} the charged lepton masses in its Fourier coefficients. 
There is no free parameter in this case except for the entire normalization which is 
{of course} scale dependent, unlike the mass ratios~\cite{ref3}.

The corresponding experimental data are in appendix~\ref{sec:app}:
the central values of $\mu$ and $\tau$ masses are
$(m_\mu, m_\tau) = (105.7, 1776)$ MeV.
Deviations of our results are at most $12.32{\%}$, so we may say that 
our calculations reproduce the experimental data well.
In the following, we mainly use the central values of the experimental results,
i.e., we neglect the errors just for simplicity.

\subsection{Case of $k=2$}
\label{sec:2.2}

In this case, we have the modular form
\be
J(q,r)=G_4+a_4G_2^2.
\ee
For the time being we ignore the second term (i.e., put $a_4=0$). 
Then we have, for the three generation case,
\be
G_4 \to M_3
= \begin{pmatrix}
11&0&0\\
4320&480&0\\
280800&165120&35040
\end{pmatrix}.
\ee
The normalized and diagonalized mass matrix becomes
\be
\left(\sqrt{M_3 M_3^T}\right)_\text{normalized}=\begin{pmatrix}
0.000163&0&0\\
0&0.964&0\\
0&0&173
\end{pmatrix}.
\ee
Here we used top quark mass of 173 GeV as the input mass.
Then the charm quark mass is obtained as 0.964 GeV,
which is a little smaller than the actual mass 1.27 GeV (by 24.1{\%}).
The up quark mass turns out to be 0.163 MeV, which is too small compared with the QCD calculations. 
We have one complex parameter $a_4$ in this case and we must work out its effect:
The detailed fit to the quark masses and also CKM matrix will be given in appendix~\ref{sec:app}.
This discussion justifies that the modular form of $k=2$ (level 4) writes down the charge $+2/3$ quark masses in its Fourier coefficients.

\subsection{Case of $k=3$}
\label{sec:2.3}

In this case, we have
\be
J(q,r)=G_6+a_6G_2^3+b_6G_2G_4.
\ee
Suppose {for the sake of argument} we take 
\be
J(q,r) = H_6=\frac{361G_2^3-G_6-50976G_2H_4}{224640},
\ee
then we 
{find}
\be
H_6\to M_3
=\begin{pmatrix}
0&0&0\\
1&0&0\\
12&-16&-2
\end{pmatrix}.
\ee
We regard the modular form of $k=3$ as an expression of the charge $-1/3$ quark masses.
{With the} QCD calculated bottom quark mass of 4.18 GeV as an input mass,
{we obtain}
\be
\left(\sqrt{M_3 M_3^T}\right)_\text{normalized}=\begin{pmatrix}
0&0&0\\
0&0.167&0\\
0&0&4.18
\end{pmatrix}.
\ee
The down quark mass is zero and the strange/bottom mass ratio is 
{off by a} factor of $1.6$.
Of course we have two complex parameters $a_6,b_6$ to be fixed in this case, and we must adjust these parameters to get more precise fit to the experimental data. 

As shown above, in the case of $k=1$ where there is no adjustable parameter the fit is almost perfect, and the other two cases require refinement but it is amazing that the values obtained in these cases also are not that distant from the experimental data. 
Now we need to choose appropriate values for $a_4, a_6$ and $b_6$.
In fact, these complex parameters are needed to fit the CKM matrix which contains some phase factor to explain the CP violation. 
See appendix~\ref{sec:app} for the concrete calculation.

\subsection{Case of $k=4$}
\label{sec:2.4}

In this case, we assume the modular form describes the neutrino masses.
The neutrino has two possibilities: 1. Dirac neutrino and 2. Majorana neutrino.

In the case of pure Dirac neutrino, 
the mass matrix becomes
\be\label{eq28}
M_D = a_8G_2^4+b_8G_4^2+c_8G_4G_2^2+G_6G_2
\ee
where
\be
G_2^4 &=& \begin{pmatrix}
1&0&0\\
576&192&0\\
154944&84480&15168
\end{pmatrix},\quad
G_4^2 = \begin{pmatrix}
121&0&0\\
95040&10560&0\\
25300800&7779840&1001280
\end{pmatrix} \nt
G_4G_2^2 &=& \begin{pmatrix}
11&0&0\\
7488&1536&0\\
1911744&878592&113856
\end{pmatrix},\quad
G_6G_2 = \begin{pmatrix}
361&0&0\\
85248&18336&0\\
39242304&18822912&1235136
\end{pmatrix}.
\ee

In the case of Majorana neutrino with seesaw approximation,
the mass matrix is given as
\be\label{eq29}
M_M = M_D M_R^{-1} M_D^T.
\ee
Although the right-handed Majorana mass $M_R$ has the same form as in equation (\ref{eq28}), it turns out that it has the following unique form 
since it must be a symmetric matrix, 
\be\label{eq29a}
M_R = d_8 \begin{pmatrix}
1&0&0\\
0&-720&0\\
0&0&-82080
\end{pmatrix}.
\ee
In this work we discuss these two limiting cases:
one is the pure Dirac case corresponding to Majorana mass $= 0$ 
and the other is the seesaw case where Majorana mass is much larger than Dirac mass. 
The actual data fitting is done in appendix~\ref{sec:app}. 

\subsection{Case of $k\geq 5$}

For $k=5$, for example, we have the modular forms 
$G_2^5, G_2^3G_4, G_2G_4^2, G_2^2G_6, G_4G_6$.
This sort of new flavor particles presumably has 
{neither} charges nor color charges, but they may have some week interactions in addition to gravitational interactions. 
Therefore, they 
{may} be a good candidate for the dark matter.

\section{Some additional considerations}

\subsection{Lagrangian}

We may write down the kinetic energy part of the Lagrangian using the fields defined in equation (\ref{eq16}).
We have
\be
K_R(z) &=&
\sum_{i,j=-\infty}^\infty \overline\psi^{a,\alpha}_{Rj}\left(D^{b,\beta}_{a,\alpha}\right)_\mu \gamma^\mu \psi_{b,\beta,Ri}e^{-2\pi iz}e^{2\pi jz} \\
K_L(z') &=& 
\sum_{i,j=-\infty}^\infty \overline\psi^{a,\alpha}_{Lj}\left(D^{b,\beta}_{a,\alpha}\right)_\mu \gamma^\mu \psi_{b,\beta,Li}e^{-2\pi iz'}e^{2\pi jz'}
\ee
where the indices $a, b$ indicate flavor type and $\alpha,\beta$ are indices for the gauge group representation. 
The right and left modes can belong to different representations.
The covariant derivative includes the gauge field $A_\mu$:
\be
\left(D^{b,\beta}_{a,\alpha}\right)_\mu = i\delta^b_a\delta^\beta_\alpha\partial_\mu
+\left(A^{b,\beta}_{a,\alpha}\right)_\mu.
\ee
Then the kinetic part of the Lagrangian density is given by
\be
\int K_R(z) dz + \int K_L(z') dz'.
\ee
To 
{maintain} the modular invariance we must impose the modular transformation:
\be
K_R(z) ~\to~ (\gamma z+\delta)^2K_R(z),\quad
K_L(z') ~\to~ (\gamma' z'+\delta')^2K_L(z'),
\ee
which means the kinetic term is a single variable modular form of level 2 in contrast to the Yukawa coupling.

\subsection{Supersymmetrization}

We may trivially write the Lagrangian in a supersymmetric 
{form}.
Corresponding to equation (\ref{eq9}), we define
\be
F(q^{-1},r^{-1})=gH\sum_{i,j=0}^{G-1}\Phi_{Rj}\Phi_{Li}q^{-i}r^{-j}.
\ee
Corresponding to equation (\ref{eq12}), we 
obtain
\be\label{Ysusy}
Y=\sum_{i,j=0}^{G-1}g_{ij}\Phi_{Rj}\Phi_{Li}H\big|_{\theta\theta}
=g\sum_{i,j=0}^{G-1}U_{Lik}^\dagger\lambda_kU_{Rkj}\Phi_{Rj}\Phi_{Li}H\big|_{\theta\theta}
\ee
where $\Phi_{Rj}$ and $\Phi_{Li}$ are the chiral fields corresponding to a certain flavor. 
Then we have
\be
g_{ij}=gU_{Lik}^\dagger \lambda_k U_{Rkj}
\ee
for $i,j=0,1,\ldots,G-1$.
Using a standard form for the chiral field 
$\Phi=A+\sqrt{2}\theta\psi+\theta\theta F$~\cite{ref5}, we get
\be
\Phi_{Rj}\Phi_{Li}H\big|_{\theta\theta}
=\left(F_{Rj}A_{Li}+A_{Rj}F_{Li}\right)H+A_{Rj}A_{Li}F_H
-\left(A_{Rj}\psi_{Li}-\psi_{Rj} A_{Li}\right)\psi_H-\psi_{Rj}\psi_{Li} H. \nt
\ee
Then the Yukawa coupling (\ref{Ysusy}) can be written as
\be
Y=g\sum_{k=0}^{G-1}\lambda_k\left[
\chi_{Lk}\chi_{Rk}H +
\left(B_{Lk}\chi_{Rk}-\chi_{Lk}B_{Rk}\right)\psi_H
+\left(G_{Lk}B_{Rk}+B_{Lk}G_{Rk}\right)H
+B_{Lk}B_{Rk}F_H
\right]\nt
\ee
where
\be
\chi_{Lk}&=&U^\dagger_{Lik}\psi_{Li},\quad \chi_{Rk}=\psi_{Rj}U_{Rjk},\nt
B_{Lk}&=&U^\dagger_{Lik}A_{Li},\quad B_{Rk}=A_{Rj}U_{Rjk},\nt
G_{Lk}&=&U^\dagger_{Lik}F_{Li},\quad G_{Rk}=F_{Rj}U_{Rjk}.
\ee
The kinetic energy part is given by
\be
K_R &=& \sum_{i=0}^\infty \Phi^\dagger_{Ri}\Phi_{Ri}\big|_{\theta\theta\overline\theta\overline\theta},\nt
K_L &=& \sum_{i=0}^\infty \Phi^\dagger_{Li}\Phi_{Li}\big|_{\theta\theta\overline\theta\overline\theta},\nt
K_H &=& \sum_{i=0}^\infty \Phi^\dagger_{H}\Phi_{H}\big|_{\theta\theta\overline\theta\overline\theta},
\ee
with
\be
\Phi^\dagger_{Ri}\Phi_{Ri}\big|_{\theta\theta\overline\theta\overline\theta}
=
G^\dagger_{Ri}G_{Ri} + B_{Ri}^\dagger \Box B_{Ri} 
+i\partial_m\overline\chi_{Ri}\overline\sigma^m\chi_{Ri}
\ee
and the similar forms for $\Phi_{Li}, \Phi_H$.

\section{Calculation of the geometry of the moduli space of Calabi-Yau manifold}
\label{sec:4}

In superstring theory, the generation number is customarily explained as the number of zero modes determined by a topological quantity. 
Our approach is different from this interpretation as explained in Introduction. 
We integrate over the modular variables when we define the low energy Lagrangian 
{that includes} Yukawa couplings among Higgs and fermions. 
Identifying the modular variables with the Calabi-Yau moduli, this means that we superpose vacuum states defined by each modulus. 
Yukawa couplings resolve the degeneracy of the vacua and each vacuum is defined by the number of generations $G$. 
At low energy we know that $G=3$,
but there may be phase transitions when we go to high energy. 
At the highest energy we may even {reach} $G\to\infty$.

Another 
{observation if} we want to interpret our result in the 
{context of} string theory is that our case may not be consistent with the grand unification. 
In fact, each flavor 
corresponds to a modular form of different level: level 2 for charged leptons, level 4 for $+2/3$ quarks, level 6 for $-1/3$ quarks 
and level 8 for neutrinos.
It is not entirely excluded that it is consistent with the grand unification, because we may have finite number of generation $G$ even at grand unified scale, and we may not worry about maintaining the modular invariance anyway.

With these conceptual modifications, our Yukawa coupling before the modular variable integration may be interpreted as coming from the compactification of the superstring theory. 

First, we assume that the following formula derived first by Strominger and Witten~\cite{ref6} is correct in spite of above conceptual modifications:
\be\label{eq41}
J(q,r,w)=J(q,r)J_H(w)
=\frac{1}{g}\sum_{i,j,k}g_{ij}h_k q^ir^jw^k
=\int_K a^\mu\wedge b^\nu\wedge c^\rho\wedge \Omega_{\mu\nu\rho}.
\ee
where $K$ is a certain Calabi-Yau Manifold and $\Omega$ is a holomorphic 3-form. 
The $a,b,c$ originate from gauge fields (principal or vector bundle) in the compactified Calabi-Yau space and {are} interpreted as harmonic (massless) $(0,1)$-form. 
If we restrict to the case of moduli corresponding to the complex structure deformation, rather than the K\"{a}hler structure deformation, the $(0,1)$-form $a,b,c$ must originate in the $(2,1)$-form.
The gauge group $A$ is the maximal subgroup such that, for example,
\be
E_8\otimes E_8 \supset A\otimes SU(3)\otimes SU(2)\otimes U(1).
\ee
We restrict ourselves to this case, and then it is shown by Candelas and de la Ossa~\cite{ref7} that the rightmost hand side of equation (\ref{eq41}) can be written as
\be
\int_K a^{\alpha\mu}\wedge b^{\beta\nu}\wedge c^{\gamma\rho}\wedge \Omega_{\mu\nu\rho}
=\frac{\partial^3 \cG}{\partial z^\alpha \partial z^\beta \partial z^\gamma}.
\ee
Here the moduli variables $z^\alpha$ ($\alpha=1,2,\ldots,\text{Betti number }b_{2,1}$) 
are chosen to be the periods themselves:
\be
z^\alpha = \int_{A^\alpha} \Omega
\ee
where $A^\alpha$ is an appropriate homology basis.

By identifying our modular variables with the 
complex structure variables $z^\alpha$~\cite{ref7}, we can explicitly calculate $\cG$ and, therefore, 
the K\"{a}hler potential $K$ is
\be
e^K = -i\left(
z^\alpha\frac{\partial}{\partial\overline z^\alpha}\overline \cG
-\overline z^\alpha\frac{\partial}{\partial z^\alpha}\cG
\right)
\ee
and the K\"{a}hler metric of the moduli space of the Calabi-Yau manifold is
\be
G_{\alpha\beta}=\frac{\partial^2}{\partial z^\alpha \partial z^\beta}\cG.
\ee
The precise relation between our modular variables $q,r$ and $w$ and the period $z^\alpha$ must respect the scaling behavior under $z\to\lambda z$:
\be
\cG(\lambda z) = \lambda^2\cG(z),\quad
\frac{\partial^3 \cG}{\partial z^\alpha \partial z^\beta \partial z^\gamma}
\to
\lambda^{-1}
\frac{\partial^3 \cG}{\partial z^\alpha \partial z^\beta \partial z^\gamma},
\ee
whereas the scaling behavior of a modular form depends on its level.
Here we consider the $\SL(2,\bbZ(\sqrt{2}))$ transformation (\ref{eq4})
with $\beta,\gamma,\beta',\gamma'=0$:
\be
z\to \frac{\alpha}{\delta} z = \alpha^2 z,\quad
z'\to \frac{\alpha'}{\delta'} z' = \alpha'{}^2 z'.
\ee
With $q=e^{2\pi i\tau}, r=e^{2\pi i\sigma}$ and $w=e^{2\pi i\rho}$, we have
\be
\tau\to\alpha^2\tau,\quad
\sigma\to\alpha^2\sigma,
\ee
since $\alpha$ must be equal to $\alpha'$ so that $\tau$ and $\sigma$ have the same scaling factor. 
This means that the $\sqrt{2}$ term in $\alpha = a+b\sqrt{2}$ must be zero and the scaling is guaranteed only for integers.
If one allows this, then we obtain 
\be
J(q,r)\to\left(\delta\delta'\right)^{2k}J(q,r) = \left(\alpha\alpha'\right)^{-2k}J(q,r)
=\alpha^{-4k} J(q,r)
\ee
and
\be
\rho\to\alpha^2\rho,\quad
J(w)\to\alpha^{-h}J(w)
\ee
where $h$ is the level of the Higgs modular form.
Therefore,
\be
J(q,r)J(w)\to
\alpha^{-h-4k}J(q,r)J(w),
\ee
then we can put
\be
\alpha^{-h-4k} =: \lambda^{-1}
\ee
and
\be
\tau\to
\lambda^\frac{2}{h+4k}\tau,\quad
\sigma\to\lambda^\frac{2}{h+4k}\sigma,\quad
\rho\to\lambda^\frac{2}{h+4k}\rho.
\ee
This shows that the period variables $z^\alpha$ are given by our modular variables:
\be\label{eq54}
z^\alpha=\left(\tau^\frac{h+4k}{2},\sigma^\frac{h+4k}{2},\rho^\frac{h+4k}{2}\right).
\ee
There are four of these combinations corresponding to 
charged leptons ($k = 1$), charge $+2/3$ quarks ($k = 2$), charge $-1/3$ quarks ($k=3$),
and neutrinos ($k=4$):
\be
\cG=\sum_{f=1}^4 \cG_f.
\ee
Although $\rho$ corresponds to the Higgs field, each combination has a different relation 
{between} $\rho$ and $z^\alpha$ as in equation (\ref{eq54}) because each combination has its own value of $k$. This means that there are 
{multiple} modular variables corresponding to the Higgs particle, which is acceptable because these variables 
{turn out} just {to be} integration variables. 
We obtain 
\be
J(q,r,w)&=&J(q,r)J_H(w)=\frac{1}{g}\sum_{i,j,k}g_{ij}h_kq^ir^jw^k \nt
&=&\frac{\partial^3\cG_f}{\partial z^\alpha\partial z^\beta\partial z^\gamma}
=\frac{\left(\frac{2}{h+4k}\right)^3}{\sqrt{(\tau\sigma\rho)^{h+4k-2}}}
\frac{\partial^3\cG_f}{\partial\tau\partial\sigma\partial\rho}.
\ee
Therefore,
\be
\cG_f=\left(\frac{h+4k}{2}\right)^3
\int^{\tau_f}\int^{\sigma_f}\int^{\rho_f}\sqrt{(\tau\sigma\rho)^{h+4k-2}}
J(q,r)J_H(w)d\tau d\sigma d\rho.
\ee
Then the metric of the moduli space of the Calabi-Yau manifold is given by
\be\label{g_moduli}
G_{\alpha\beta}=\sum_{f=1}^4G_{\alpha\beta,f}
=\sum_{f=1}^4\frac{\partial^2}{\partial z^\alpha\partial z^\beta}\cG_f.
\ee
For example,
\be
G_{\tau\sigma,f}
=\frac{\left(\frac{2}{h+4k}\right)^2}{\sqrt{(\tau_f\sigma_f)^{h+4k-2}}}
\frac{\partial^2\cG_f}{\partial\tau_f\partial\sigma_f}
=\frac{h+4k}{2}J(q_f,r_f)\int^{\rho_f}\sqrt{\rho_f^{h+4k-2}}J_H(w)d\rho.
\ee
We remark that the other derivatives such as
$
\frac{\partial^3\cG_f}{\partial\tau^3},
\frac{\partial^3\cG_f}{\partial\tau^2\partial\sigma},
$
etc.
can correspond to some Yukawa couplings, but all these seem not to appear in physics because of the gauge symmetry of the theory. 
For example, 
$\frac{\partial^3\cG_f}{\partial\tau^3}$
{could} potentially correspond to the triple Higgs coupling, but it is forbidden by the standard model symmetry.

The K\"{a}hler metric of the moduli space (\ref{g_moduli}) is related to the Calabi-Yau metric through the equation:
\be\label{eq60}
G_{\alpha\beta} = \frac{1}{2V}\int_M
g^{\kappa\overline\mu} g^{\lambda\overline\nu}\left(
\frac{\partial g_{\kappa\lambda}}{\partial z^\alpha}
\frac{\partial g_{\overline\mu\overline\nu}}{\partial z^\beta}
+
\frac{\partial B_{\kappa\lambda}}{\partial z^\alpha}
\frac{\partial B_{\overline\mu\overline\nu}}{\partial z^\beta}
\right) d^6x
\ee
where $g_{\mu\nu}$ is the Calabi-Yau metric, $B_{\mu\nu}$ is a 2-form related to
$g_{\mu\nu}$ by supersymmetry,
and $V$ is the volume of the Calabi-Yau manifold. 
For example,
\be
G_{\tau\sigma,f} = \frac{1}{2V}
\frac{\left(\frac{2}{h+4k}\right)^2}{\sqrt{(\tau_f\sigma_f)^{h+4k-2}}}
\int_M
g^{\kappa\overline\mu} g^{\lambda\overline\nu}\left(
\frac{\partial g_{\kappa\lambda}}{\partial \tau_f}
\frac{\partial g_{\overline\mu\overline\nu}}{\partial \sigma_f}
+
\frac{\partial B_{\kappa\lambda}}{\partial \tau_f}
\frac{\partial B_{\overline\mu\overline\nu}}{\partial \sigma_f}
\right) d^6x.
\ee
If we restrict to the minimum Calabi-Yau manifold, meaning that 
all of its moduli are 
directly determined by the experiments as above, we may be able to determine its metric $g_{\mu\nu}$ by solving equation (\ref{eq60}) together with the Ricci flat and K\"{a}hler constraints for $g_{\mu\nu}$. 
We would like to come back to this issue in a future publication.

\section{Concluding remarks}

\begin{enumerate}
\item
As we have shown above, the hypothesis of flavor moonshine is at least correctly realized experimentally to some extent. 
We need to use multi-variable modular forms for this purpose. 
These forms are well studied in mathematics as a brunch of number theory and they constitute a part of more general forms called Hilbert modular form~\cite{ref8}.

\item
We use only the Fourier coefficients of these forms to define the Yukawa coupling and the modular invariance of the total Lagrangian is assumed~\cite{ref9}. 
As such, it corresponds to the procedure of integrating over the modular variables which are identified as Calabi-Yau moduli if we combine our model 
{with} string theory. 
We do not regard these moduli as scalar fields to be stabilized. 
{Insofar as we can see}, there seems to be no justification 
{for regarding} them as scalar fields. 
Therefore, our treatment of 
{them as} moduli to be integrated out when we define the low energy action seems to be a natural process.

\item
Of course, there are many mysteries to be solved. 
Why nature seems to choose a very specific form such as the one we used 
{that} is based on $\SL(2, \bbZ(\sqrt{2}))$? 
Why $k=1$ for charged leptons, $k=2$ for charge $+2/3$ quarks, $k=3$ for charge $-1/3$ quarks, and $k=4$ for neutrinos?

There remain a lot of works to be done: 
How good or bad are the other modular groups like $\SL(2, \bbZ(\sqrt{N
}), \SL(2, \bbZ(i))$ etc.? 
Can we extend the modular form to be 
{more} than two variables? 
What exactly is the mathematical moonshine for the modular form of two variables? 
If we understand the mathematical implication of the matrices which appear in the Fourier coefficients of two variable modular forms, we will be able to prove the flavor moonshine by understanding the physical principle 
{that identifies} mass matrices with these matrices.

\item
Probably more urgent work from the string theory standpoint is to find out the specific Calabi-Yau metric by solving equation (\ref{eq60}) and 
{to elucidate} its other physical consequences. 
{Further} questions {arise such} 
as: 
Do we have a grand unified scale? 
Do we have a phase transition from $G=3$ to $G\geq 4$ at some point in higher energy? 

\item
Experimentally, we need to explore the property of Higgs particle in more detail,
especially its coupling to low mass particles such as $u, d, e, \mu$ and even neutrinos. 
Construction of ILC, therefore, is urgent. 
A good neutrino facility is also highly desirable. 
The Higgs particle is indeed the ``God particle," the term coined by Leon Lederman~\cite{ref10}, in the sense that its Yukawa couplings determine the highest energy physics without 
{the need to perform} the highest energy experiments. 

\item
It is possible that the whole idea of flavor moonshine is just nonsense~\cite{ref11},
although the agreement with the experimental data seems to us 
too good to be just an accident.

\end{enumerate}

\appendix
\section{Numerical fitting for experimantal data}
\label{sec:app}

We calculated numerically the CKM and PMNS matrices and fit the experimental data 
{to} them. 
In the former case we have three complex parameters $a_4,a_6$ and $b_6$ as shown in equations (\ref{eq7}) and (\ref{eq8}). 
For the PMNS matrix we have two choices of pure Dirac neutrino or Majorana neutrino (with seesaw approximation).
In either way, we have again three complex parameters $a_8,b_8$ and $c_8$ shown in equation (\ref{eq28}).
Since the parameter $d_8$ in equation (\ref{eq29a}) is an overall factor,
we 
need {not} consider it in our discussion.

Let us briefly explain how we get the CKM matrix,
which is parallel 
{to} the PMNS matrix.
Now we have the mass matrix $M_3$ for $u,c,t$ quarks, as in section \ref{sec:2.2},
with the complex parameters.

First we calculate the squared mass matrix as in equation (\ref{eq:gg}):
$M_3 M_3^\dagger$ or $M_3^\dagger M_3$.
Here we have the two choices 
{that} give us the same eigenvalues
but different eigenvectors.
To obtain its eigenvalues and eigenvectors, we 
{compute}
\be\label{eigen}
U^\dagger (M_3 M_3^\dagger) U = D
\quad\text{or}\quad
U^\dagger (M_3^\dagger M_3) U = D
\ee
where $U$ is a unitary matrix and $D$ is a diagonal matrix.
The masses of $u,c,t$ quarks are given by the square root of the eigenvalues:
\be
D = \begin{pmatrix}
m_u^2 & 0 & 0\\
0 & m_c^2 & 0\\
0 & 0 & m_t^2
\end{pmatrix}.
\ee
Here we have swapped the columns of $U$ and $D$ so that $m_u<m_c<m_t$.
Then the eigenvectors are regarded as the quark mass states
\be
\begin{pmatrix} u&c&t \end{pmatrix}_\text{mass}
=U
\begin{pmatrix} u&c&t \end{pmatrix}_\text{current}
=U
\ee
where we set the quark current states as
\be
u_\text{current} = \begin{pmatrix} 1\\0\\0 \end{pmatrix},\quad
c_\text{current} = \begin{pmatrix} 0\\1\\0 \end{pmatrix},\quad
t_\text{current} = \begin{pmatrix} 0\\0\\1 \end{pmatrix}.
\ee
We repeat similar calculations for $d,s,b$ quarks (see section \ref{sec:2.3}) and 
obtain
\be
\begin{pmatrix} d&s&b \end{pmatrix}_\text{mass}
=V
\begin{pmatrix} d&s&b \end{pmatrix}_\text{current}
=V
\ee
where $V$ is a unitary matrix including the eigenvectors 
of the squared mass matrix for $d,s,b$ quarks.
Note that, by definition, the current quarks should satisfy
\be
\begin{pmatrix} u^\dagger\\c^\dagger\\t^\dagger \end{pmatrix}_\text{current}
\begin{pmatrix} d&s&b \end{pmatrix}_\text{current}
= I. 
\ee
Therefore, the CKM matrix can be calculated as
\be
\text{CKM} = 
\begin{pmatrix} 
u^\dagger d & u^\dagger s & u^\dagger b\\
c^\dagger d & c^\dagger s & c^\dagger b\\
t^\dagger d & t^\dagger s & t^\dagger b
\end{pmatrix}_\text{mass}
= U^\dagger V.
\ee

For calculation of the PMNS matrix, we use
the mass matrix $M_3$ for charged leptons in section \ref{sec:2.1}
and $M_3=M_D~\text{or}~M_M$ for neutrino in section \ref{sec:2.4}.

\subsection{Methods}

Our goal is to find a set of complex parameters 
{that best} fit the experimental results. 
The experimental results we use here are
\begin{itemize}
\item 
the absolute values of the elements of the mixing (CKM or PMNS) matrix $\zeta_{ij}$
\item
the ratios of masses $\xi_k$.
\end{itemize}
The mixing matrices in both cases have $3\times 3=9$ elements.
Note that the CP violation phases are not used for our fittings.
For quark masses, we choose the parameters $\xi_k=(m_t/m_c, m_b/m_s)$.
This means we do not fit $u$ and $d$ quark masses:
In all the results we 
{obtained} they are much smaller than experimental results,
just as we saw in section \ref{sec:2.2}.
For lepton masses, we choose
$\xi_k = \Delta m_{21}^2/\Delta m_{32}^2$,
i.e., a ratio of difference of squared neutrino masses.
Since the masses of $e,\mu$ and $\tau$ are already fixed, as in section \ref{sec:2.1}, 
we have no parameters 
{to fit} them.

Then we define the loss function to measure a ``difference'' between our results and the experimental results:
\be\label{loss}
\text{Loss} = 
\sum_{i,j=1}^3 \left| \log \frac{\zeta_{ij}^\text{cal}}{\zeta_{ij}^\text{exp}} \right|
+2 \sum_k \left| \log \frac{\xi_k^\text{cal}}{\xi_k^\text{exp}} \right|
\ee
where $\zeta_{ij}^\text{exp}$ and $\xi_k^\text{exp}$ are the experimental results,
while $\zeta_{ij}^\text{cal}$ and $\xi_k^\text{cal}$ are our results of numerical calculations
(which depend on the three complex parameters).
The factor of 2 exists in the second term
{ensures that} the contribution from this term cannot be much smaller than 
{that from} the first term:
the ratios of masses have only 2 (or 1) parameters in the quark (or lepton) case,
while the mixing matrix has 9 parameters.

Now let us search the complex parameters at the minimum of the loss function (\ref{loss}).
First we divide the 3 complex parameters into 6 real parameters $x_i$.
Since the following discussion includes calculating eigenvectors of matrices,
the iterative approximation with gradient descent is not suitable to be used.
Instead, we choose 11 lattice points for each real parameter 
\be
x_i = x_i^0-5\delta x,~ x_i^0-4\delta x,~ \ldots,~ x_i^0+5\delta x.
\ee
where {for simplicity} the lattice spacing $\delta x$ is the same for all $i$, 
and at first we set $x_i^0=0$ for all $i$.
Then we have $11^6$ lattice sites in total.

After calculating the loss function (\ref{loss}) at all the lattice sites,
we find a set of parameters $x_i^\text{min}$ with the minimum loss among them.
Next we set $x_i^0 = x_i^\text{min}$ and $\delta x\to \delta x/6$,
and repeat this procedure in six times.
Finally the lattice spacing becomes $\delta x/6^6$.

We tried several cases satisfying $10^{-3} \leq \delta x/6^6\leq 10^{-2}$,
and calculated both cases of the squared mass matrix (\ref{eigen}).
Then we obtain a certain set of parameters with the minimum loss among all the results we obtained.
In our discussion we regard it as the best fit for the experimental results.

\subsection{CKM matrix}

The best fit we obtained for the CKM matrix is
\be
\text{CKM}=\begin{pmatrix}
0.974&0.226&0.004 e^{-1.17i}\\
-0.226&0.973&0.043\\
0.009 e^{-0.435i}&-0.042&0.999
\end{pmatrix}
\ee
with quark masses
\be
\left(m_u,m_c,m_t\right)&=&(5.30\times 10^{-5},1.30,173) ~\text{GeV} \nt
\left(m_d,m_s,m_b\right)&=&(1.18\times 10^{-6},0.013,4.18) ~\text{GeV}.
\ee
Here we input $m_t$ and $m_b$ for normalization.
The CKM can be expressed in terms of Wolfenstein parameters
\be
\begin{pmatrix}
1-\frac{\lambda^2}{2} & \lambda & A\lambda^3(\rho-i\eta) \\
-\lambda & 1-\frac{\lambda^2}{2} & A\lambda^2\\
A\lambda^3(1-\rho-i\eta) & -A\lambda^2 & 1
\end{pmatrix}
+\cO(\lambda^4),
\ee
and we obtain
\be
\lambda=0.226,~ A=0.839,~ \rho=0.161,~ \eta=0.382.
\ee
The experimental values for these are~\cite{PDG} 
\be
\lambda=0.226,~ A=0.836,~ \rho=0.125,~ \eta=0.364
\ee
with quark masses
\be
\left(m_u,m_c,m_t\right)&=&(2.2\times 10^{-3},1.27,173) ~\text{GeV} \nt
\left(m_d,m_s,m_b\right)&=&(4.7\times 10^{-3},0.093,4.18) ~\text{GeV}.
\ee
Note that, again, we look at only the central values of the experimental data.

Some comments are in order for these results:
\begin{enumerate}
\item
The agreement is generally excellent.

\item
Masses of $u,d,s$ quarks come out to be rather small. 
This is due to large hierarchical property of the mass matrices. 
Lattice QCD mass is somewhat different from the Higgs coupling, especially its renormalization corrections,
but it is not clear at this time whether this fact can account for the difference.

\item
The CKM matrix has also renormalization corrections~\cite{ref12}. 
The fact that our result is not far from the experimental value may indicate that our theory is indeed a low energy theory rather than the very short distance theory. 

\end{enumerate}

\subsection{PMNS matrix}

{Our} best fit 
for the PMNS matrix is {obtained} as follows.
We discuss the two cases of pure Dirac neutrino and Majorana neutrino with seesaw approximation.
In each case, neutrino masses can be in the normal order ($m_1<m_2<m_3$) or the inverted order ($m_3<m_1<m_2$).

\subsubsection{Case of pure Dirac neutrino}

When neutrino masses are in the normal order, the best fit is 
\be
\text{PMNS} = \begin{pmatrix}
0.919 & 0.183 & 0.349 e^{-1.49i} \\
0.304 e^{2.05i} & 0.598 e^{0.09i} & 0.742 \\
0.250 e^{0.98i} & 0.780 e^{3.09i} & 0.573 
\end{pmatrix} 
\ee
with neutrino mass differences
\be
\left(\Delta m_{21}^2, \Delta m_{32}^2\right)
=\left(m_2^2-m_1^2, m_3^2-m_2^2\right)
=(7.53\times 10^{-5},3.32\times 10^{-1})~\text{eV}^2.
\ee
Here $\Delta m_{21}^2$ is our input for normalization, which is the same for all the fittings below. 

If neutrino masses are in the inverted order, the best fit becomes
\be
\text{PMNS} = \begin{pmatrix}
0.586 & 0.483 & 0.651 e^{-2.79i} \\
0.150 e^{2.12i} & 0.814 e^{0.13i} & 0.561 \\
0.796 e^{0.15i} & 0.323 e^{2.84i} & 0.511 
\end{pmatrix} 
\ee
with neutrino mass differences
\be
\left(\Delta m_{21}^2, \Delta m_{32}^2\right)
=(7.53\times 10^{-5},-7.53\times 10^{-5})~\text{eV}^2.
\ee

The PMNS matrix is can be written as
\be\label{eq:PMNS}
\begin{pmatrix}
c_{12}c_{13} & s_{12}c_{13} & s_{13} e^{-i\delta} \\
-s_{12}c_{23}-c_{12}s_{23}s_{13}e^{i\delta} & c_{12}c_{23}-s_{12}s_{23}s_{13}e^{i\delta} & s_{23}c_{13} \\
s_{12}s_{23}-c_{12}c_{23}s_{13}e^{i\delta} & -c_{12}s_{23}-s_{12}c_{23}s_{13}e^{i\delta} & c_{23}c_{13}
\end{pmatrix}
\ee
where $c_{12} = \cos\theta_{12}, s_{12} = \sin\theta_{12}, \cdots$.
Then we get
\be
s_{12}^2&=&0.0381,~ s_{13}^2=0.122,~ s_{23}^2=0.626,~\delta=1.49 \quad \text{(normal order)}\nt
s_{12}^2&=&0.404,~ s_{13}^2=0.424,~ s_{23}^2=0.547,~\delta=2.79 \quad \text{(inverted order)}.
\ee
Lepton masses in both cases of normal and inverted orders are the same as in section \ref{sec:2.1}:
\be
\left(m_e,m_\mu,m_\tau\right)
=(0.5110,107.5,1558)~\text{MeV}.
\ee

\subsubsection{Case of Majorana neutrino with seesaw approximation}

The best fit in the normal order of neutrino masses is 
\be
\text{PMNS} = \begin{pmatrix}
0.291 & 0.753^{1.96i} & 0.590 e^{1.12i} \\
0.489 e^{-2.96i} & 0.527 e^{-2.66i} & 0.695 \\
0.822 e^{0.70i} & 0.394 e^{-1.40i} & 0.411 
\end{pmatrix} 
\ee
with neutrino mass differences
\be
\left(\Delta m_{21}^2, \Delta m_{32}^2\right)
=(7.53\times 10^{-5},2.44\times 10^{-3})~\text{eV}^2.
\ee

In the inverted order of neutrino mass, the best fit is
\be
\text{PMNS} = \begin{pmatrix}
0.294 & 0.880^{3.14i} & 0.373 e^{3.14i} \\
0.490 e^{0.00i} & 0.197 e^{3.14i} & 0.849 \\
0.821 e^{3.14i} & 0.433 e^{3.14i} & 0.373 
\end{pmatrix}
\ee
with neutrino mass differences
\be
\left(\Delta m_{21}^2, \Delta m_{32}^2\right)
=(7.53\times 10^{-5},-7.53\times 10^{-5})~\text{eV}^2.
\ee

The PMNS matrix in this case is can be written as
\be
\begin{pmatrix}
c_{12}c_{13} & s_{12}c_{13} & s_{13} e^{-i\delta} \\
-s_{12}c_{23}-c_{12}s_{23}s_{13}e^{i\delta} & c_{12}c_{23}-s_{12}s_{23}s_{13}e^{i\delta} & s_{23}c_{13} \\
s_{12}s_{23}-c_{12}c_{23}s_{13}e^{i\delta} & -c_{12}s_{23}-s_{12}c_{23}s_{13}e^{i\delta} & c_{23}c_{13}
\end{pmatrix}
\begin{pmatrix}
1&0&0\\
0&e^{i\alpha_{21}/2}&0\\
0&0&e^{i\alpha_{31}/2}
\end{pmatrix},
\ee
then we get
\be
s_{12}^2&=&0.870,~ s_{13}^2=0.348,~ s_{23}^2 =0.741 \quad \text{(normal order)} \nt
s_{12}^2&=&0.900,~ s_{13}^2=0.139,~ s_{23}^2 =0.838 \quad \text{(inverted order)}
\ee
and the CP violation phases are
\be
\delta=-1.68,~ \alpha_{21} = 3.92,~ \alpha_{31}=-1.13 \quad \text{(normal order)} \nt
\delta=0.00,~ \alpha_{21} = 0.01,~ \alpha_{31}=0.00 \quad \text{(inverted order)}
\ee
modulo $2\pi$. 
Lepton masses are the same as in the case of pure Dirac neutrino.

\subsubsection{Experimental data}

The current experimental values (its central values) of PMNS matrix are
\cite{PDG,MNS}
\be
\left|\text{PMNS}\right| = \begin{pmatrix}
0.821 & 0.550 & 0.150 \\
0.304 & 0.598 & 0.742 \\
0.483 & 0.583 & 0.654 
\end{pmatrix},
\ee
the angles in the expression (\ref{eq:PMNS}) are 
\be
s_{12}^2&=&0.307,~ s_{13}^2=0.0218,~ s_{23}^2 =\begin{cases}
0.512 & \text{(normal order)} \\
0.536 & \text{(inverted order)}
\end{cases},
\ee
and the CP violation phase is 
\be
\delta&=&1.37\pi = -1.98 \quad\text{(modulo $2\pi$)}.
\ee
The lepton masses are
\be
\left(m_e,m_\mu,m_\tau\right)
=(0.5110,105.6,1777)~\text{MeV}
\ee
and the neutrino mass differences are
\be
\left(\Delta m_{21}^2, \Delta m_{32}^2\right)
=\begin{cases}
(7.53\times 10^{-5},2.44\times 10^{-3})~\text{eV}^2 & \text{(normal order)} \\
(7.53\times 10^{-5},-2.55\times 10^{-3})~\text{eV}^2 & \text{(inverted order)}
\end{cases}.
\ee

We see that
\begin{enumerate}
\item
The agreement seems the best 
for the Majorana neutrino case in the normal order,
especially at the CP violation phase and the neutrino mass difference.


\item
In that case, 
$\sin\theta_{23}$ matches well and $\sin\theta_{12}$ agrees within a factor of 3.
However, we obtain 
too large {a} value for $\sin\theta_{13}$.
The discrepancy with the experimental data could be attributed to the renormalization effect or inadequacy of our assignment.
Further study is required.

\item
In the inverted order, we obtain no good agreements and 
the neutrino mass differences {in particular} completely fail to agree.
{Since} masses have the large hierarchical property in our calculations,
{as a consequence} $|\Delta m_{32}^2|$ never exceed $|\Delta m_{21}^2|$.

\end{enumerate}

\subsection*{Acknowledgment}
We would like to thank Professors A.~Kusenko and R.~Peccei of UCLA, where part of this work was done, for their kind hospitality.
We also appreciate Professors L.~Brink, S.~Hashimoto, S.~Iso, H.~Ooguri and P.~Ramond
for their informative comments.
We thank Professor J. Miller for his reading of the manuscript and checking English.


\begin{thebibliography}{99}
\bibitem{ref1}
S. Pakvasa and H. Sugawara, Phys. Lett. {\bf 73B} (1978) 61,\\
F. Wilczek and A. Zee, Phys. Lett. {\bf 70B} (1977) 418
[Erratum: Phys. Lett. {\bf 72B} (1978) 503],\\
G. Altarelli and F. Feruglio, Rev. Mod. Phys. {\bf 82} (2010) 2701 [arXiv:1002.0211 [hep-ph]],\\
H. Ishimori, T. Kobayashi, H. Ohki, Y. Shimizu, H. Okada and M. Tanimoto, Prog.
Theor. Phys. Suppl. {\bf 183} (2010) 1 [arXiv:1003.3552 [hep-th]],\\
D. Hernandez and A. Y. Smirnov, Phys. Rev. {\bf D86} (2012) 053014 [arXiv:1204.0445
[hep-ph]],\\
S. F. King and C. Luhn, Rept. Prog. Phys. {\bf 76} (2013) 056201 [arXiv:1301.1340 [hep-ph]],\\
S. F. King, A. Merle, S. Morisi, Y. Shimizu and M. Tanimoto, New J. Phys. {\bf 16} (2014)
045018 [arXiv:1402.4271 [hep-ph]],\\
S. F. King, Prog. Part. Nucl. Phys. {\bf 94} (2017) 217 [arXiv:1701.04413 [hep-ph]],\\
C. Hagedorn, arXiv:1705.00684 [hep-ph].

\bibitem{ref6}
A. Strominger and E. Witten, Commun. Math. Phys. {\bf 101} (1985) 341.

\bibitem{ref2}
J. H. Conway and S. P. Norton, Bull. London Math. Soc. {\bf 11} (3) (1979) 308-339,\\
I. B. Frenkel, J. Lepowsky and A. Meurman, Pure and Applied Math. {\bf 134}. Academic Press. MR 0996026 (1988),\\
R. Borcherds, Invent. Math. {\bf 109} (1992) 405-444,\\
T. Eguchi, H. Ooguri and Y. Tachikawa, Exper. Math. {\bf 20} (2011) 91-96 [arXiv:1004.0956 [hep-th]].

\bibitem{ref3}
K. G. Chetyrkin and A. R\'{e}tey, 
Nucl. Phys. {\bf B583} (2000) 3 [arXiv:hep-ph/9910332]. 

\bibitem{ref4}
H. Cohn and J. Deutsch, Math. Computation {\bf 48} 177 (1987) 139.

\bibitem{ref5}
J. Wess and J. Bagger, ``Supersymmetry and Supergravity'' (revised version), Princeton University Press, 1992.

\bibitem{ref7}
P. Candelas and X. C. de la Ossa, Nucl. Phys. {\bf B355} (1991) 455.

\bibitem{ref8}
E. Freitag, ``Hilbert Modular Forms'', Springer-Verlag, 1990.

\bibitem{ref9}
During the preparation of this paper, we encountered 
a work by J. C. Criado and F. Feruglio (arXiv:1807.01125 [hep-ph]) 
and F. Feruglio (arXiv:1706.08749 [hep-ph]). 
Their work has nothing to do with moonshine, but, since they assume the modular invariance of the low energy action, although the way they impose it is very different from ours, their Yukawa coupling depends on modular forms (not their Fourier coefficients) and it may be possible to calculate the Calabi-Yau moduli space geometry in this case, too.

\bibitem{ref10}
L. Lederman and D. Teresi, ``The God Particle", Dell Publishing, 1993.

\bibitem{ref11}
Wikipedia on ``Monstrous moonshine" at 
\verb|https://en.wikipedia.org/wiki/Monstrous_moonshine|: 
The term ``monstrous moonshine" was coined by Conway, who, when told by John McKay in the late 1970s that the coefficient of $q$ (namely 196884) was precisely one more than the degree of the smallest faithful complex representation of the monster group (namely 196883), replied that this was ``moonshine" (in the sense of being a crazy or foolish idea). Thus, the term not only refers to the monster group $M$; it also refers to the perceived craziness of the intricate relationship between $M$ and the theory of modular functions.\\
$M$ in the present work refers to ``Mass" rather than ``Monster group".

\bibitem{PDG}
M. Tanabashi et al. (Particle Data Group), Phys. Rev. {\bf D98} (2018) 030001.

\bibitem{ref12}
V. Barger, M. S. Berger and P. Ohmann, 
Phys. Rev. {\bf D47} (1993) 2038 [arXiv:hep-ph/9210260]. 

\bibitem{MNS}
I. Esteban, M. C. Gonzalez-Garcia, A. Hernandez-Cabezudo, M. Maltoni and T. Schwetz,
JHEP {\bf 01} (2019) 106 [arXiv:1811.05487 [hep-ph]].

\end{thebibliography}
\end{document}